\documentclass[a4paper,11pt]{article}
\usepackage{pos}
\usepackage{mathtools}
\usepackage{siunitx}
\usepackage{subfigure}

\title{DarkSide-20k: Next generation Direct Dark Matter searches with liquid Argon}
\manuallySeparateAuthors
\author*[a]{Ioannis Manthos} 
\author{on behalf of the  DarkSide-20k Collaboration}

\affiliation[a]{School of Physics and Astronomy, University of Birmingham, B15 2TT, Birmingham, United Kingdom}

\emailAdd{i.manthos@bham.ac.uk}

\abstract{DarkSide-20k is a next-generation dual-phase Liquid Argon Time Projection Chamber, currently under construction at the Gran Sasso National Laboratory (LNGS) in Italy. 
With a 20\,t fiducial mass of liquid Argon, DarkSide-20k will probe  
 WIMP-nucleon interactions down to cross sections equal to 10$^{-48}$\,cm$^2$ for a WIMP mass of 0.1\,TeV/c$^2$. DarkSide-20k is designed to be a nearly ``instrumental background-free'' experiment, meaning that less than 0.1 background events are expected in the WIMP search region during the 200\,tonne-year planned exposure. To achieve this, the TPC is surrounded by an inner (neutron) and outer (muon) veto, while low-radioactivity underground argon (depleted in $^{39}$Ar), is used as the inner detector (TPC and inner veto) medium. Both the TPC and the veto systems are instrumented with novel cryogenic silicon photomultipliers, capable of resolving single photoelectrons and providing the required spatial and time resolution. An overview of the DarkSide-20k experimental program is reported, with a focus on the photo-detector system construction and testing procedures for the inner veto system.
}

\FullConference{The European Physical Society Conference on High Energy Physics (EPS-HEP2023)\\
 21-25 August 2023\\
Hamburg, Germany\\}

\begin{document}
\maketitle

\section{The DarkSide-20k experiment}
Dark Matter (DM) makes up more than 5 times as much of the universe as the ordinary matter we are made of, yet its nature is unknown. This fundamental open question has
initiated one of the greatest hunts in the history of physics \citep{Cebrian:2022brv}. The main search channel aims at the detection of DM interaction with ordinary matter, exploiting liquid noble targets at large-scale experiments with unprecedentedly sensitive detectors. The Global Argon Dark Matter Collaboration designs, builds and will operate DarkSide-20k \cite{Darkside}; the largest, most ambitious experiment thus far, using argon as the target, designed to detect (or exclude) DM in the very promising mass range 1\,GeV/c$^2$ - 10\,TeV/c$^2$.
\begin{figure}[h]
\centering
\includegraphics[width=0.45\linewidth]{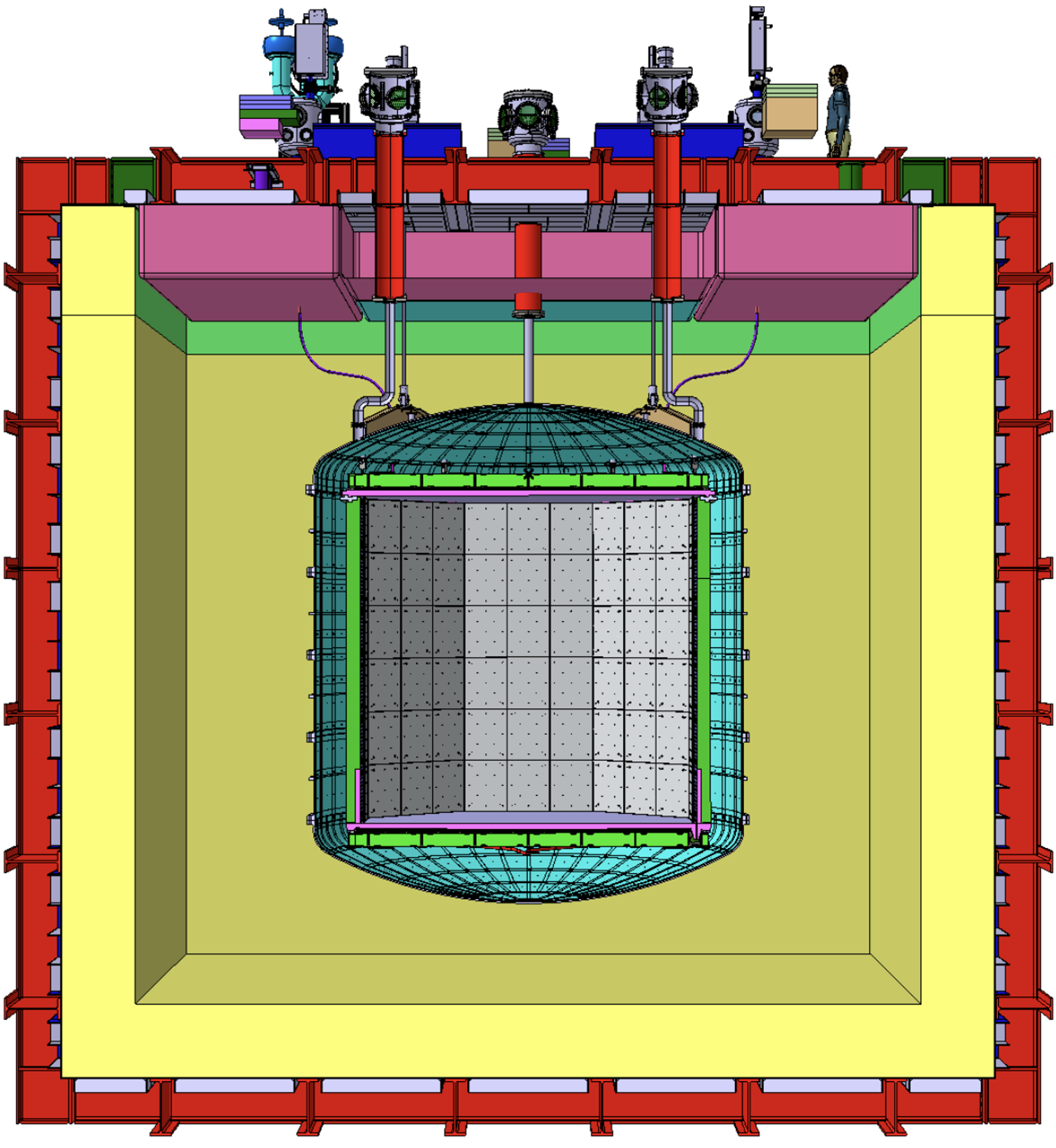}
\caption{Schematic of the DarkSide-20k detector.\label{fig:fig1}}
\end{figure}

DarkSide-20k is a massive liquid argon (LAr) detector (shown in Fig.\ref{fig:fig1}) that is currently under construction in LNGS, Italy, with the aim to start operations from 2026 for a planned 200\,tonne-year exposure. The inner detector is a time projection chamber (TPC) hosting 51 tonnes of underground argon (UAr), with more than 1400 times less background than atmospheric LAr, surrounded by a further 32\,t of UAr in the veto; TPC and veto are instrumented with 26\,m$^2$ of novel silicon photomultipliers (SiPM). This is immersed in an outer cosmic veto, consisting of a 600\,t LAr bath, sparsely instrumented with the same silicon detector technology. The inner detector employs the TPC and veto to mitigate the most critical background to DM searches, which comes from neutron scattering, whilst the outer cosmic veto detects and tags external neutrons and muons. This biggest ever constructed (up-to-date) dual phase TPC, features the asset of very strong pulse shape discrimination between electron and nuclear recoils that Ar provides, offering  an electron recoil  background rejection factor of $2.4\times10^{-8}$ for the 44-89\,keV$_{ee}$ energy range.
\section{Innovations of  DarkSide-20k for background reduction}
With the objective to be a nearly ``instrumental background free'' detector, significant R\&D has been performed towards the minimization of the instrumental background. The main topics are summarized in this section and are related with the use of underground Ar, the neutron veto and the photodetectors technology. Complemented by a deliberate material selection, extensive radiopurity assays, and exposure minimization, a background of less than 0.1 events is expected after all cuts in a full exposure of 200\,tonne-year.
\subsection{Underground Argon}
The predecessor experiment, DarkSide-50, demonstrated a depletion factor $1400 \pm 200$ \cite{PhysRevD.93.081101} in the presence of the $\beta$ emitter $^{39}$Ar isotope, by using UAr as the  target material. The same approach is followed for DarkSide-20k \cite{Pesudo}, UAr is extracted from an underground CO$_{2}$ well (Urania plant, in Cortez, CO, USA) that contains Ar free from the activated $^{39}$Ar.  UAr will be further chemically purified to detector-grade Ar in a 350\,m cryogenic distillation column (Aria facility, Sardinia, Italy), to remove non-Ar isotopes \citep{Aria}. An assessment of $^{39}$Ar presence on UAr samples will be performed with the DaRT detector \citep{Garcia_2020} at the Canfranc Underground Laboratory (LSC) in Spain.
\subsection{Neutron veto}
\begin{figure}[h]
\centering
\includegraphics[width=0.45\linewidth]{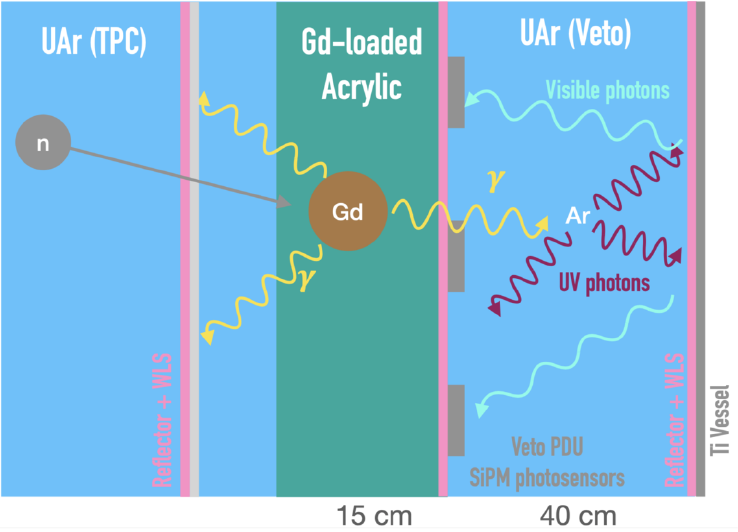}
\caption{Description of the neutron veto on DarkSide-20k inner detector.\label{fig:fig2}}
\end{figure}

Neutrons with $\mathcal{O}(\textrm{MeV})$ energy, mostly originating from decays of the radioactive contaminants of the detector material, form a nuclear recoil  background that mimics the expected WIMP interactions with the target medium. This dangerous background must be efficiently identified and mitigated. The walls of the TPC besides offering mechanical support, perform the neutron veto. This is achieved by the, 15\,cm thick, gadolinium-loaded polymethylmethacrylate (Gd-PMMA) material that they are constructed of. The rich in hydrogen PMMA efficiently thermalises the MeV fast neutrons, which subsequently are captured due to the high (n,$\gamma$) Gd neutron capture cross-section, emitting multiple  $\gamma$-rays per neutron with energies up to 8\,MeV. Those are detected by the photo detectors through the scintillation photons they induce on the LAr of the veto buffer (shown in Fig.\ref{fig:fig2}). While WIMP events will deposit energy only on the TPC fiducial volume, neutron events will be ruled out by the veto by introducing an exclusion time of 600-800\,$\si{\micro \textrm{s}}$ for WIMP detection.
\begin{figure}[h]
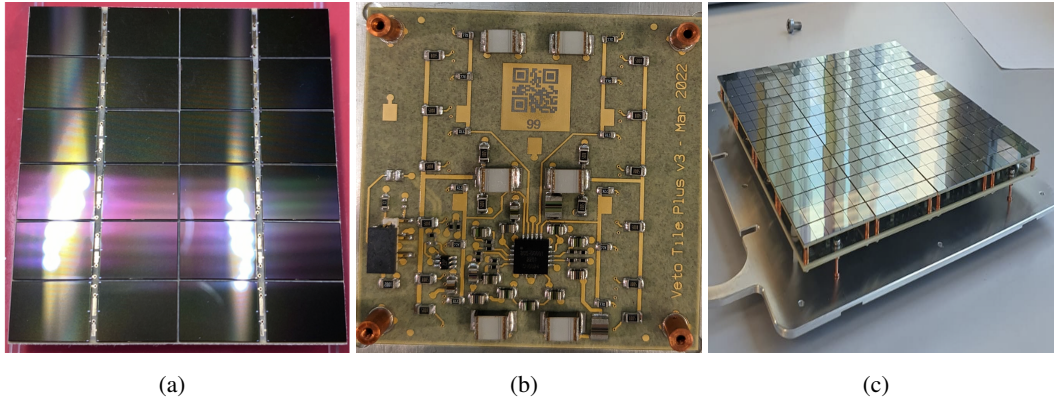

\centering
\subfigure[\label{fig:fig3a}]{\includegraphics[width=0.3\linewidth,height=11pc]{Fig3a.pdf}}
\subfigure[\label{fig:fig3b}]{\includegraphics[width=0.3\linewidth,height=11pc]{Fig3b.pdf}}
\subfigure[\label{fig:fig3c}]{\includegraphics[width=0.3\linewidth,height=11pc]{Fig3c.pdf}}
\caption{\subref{fig:fig3a} The front \subref{fig:fig3b} and the back side of the veto Photo Detection Module. \subref{fig:fig3c} The completed veto Photo Detection Unit.\label{fig3}}
\end{figure}
\subsection{Photodetectors technology}
Accurate light collection is of paramount importance for the detector's performance. For this purpose, the DarkSide-20k collaboration choose the silicon-photomultipliers (SiPM) photodetectors technology \cite{simp}, and performed an intense R\&D program. A SiPM is comprised of Single Photon Avalanche Diode (SPAD) semiconductor devices based on a p-n junction, operating independently with reverse bias well above breakdown voltage (Geiger mode). With a typical size of 25-30\,$\si{\micro \textrm{m}}$, for a SiPM 94.900 SPADs are placed in an array, forming a surface of 8$\times$12\,mm. 

SiPM can achieve an exceptional charge resolution, $\mathcal{O}(1\%)$, an improvement of one order of magnitude compared to  photomultiplier (PMT) detectors. The photon detection efficiency, which includes both the quantum efficiency and the SPAD fill factor ($>90\%$), is $\approx45\%$ on the full detector surface. Furthermore, SiPMs expose more than 10 times lower radioactivity/cm$^2$ than PMTs. The DarkSide-20k SiPMs perform with a breakdown voltage of 28\,V, signal to noise ratio >8 and dark current rate <0.001\,Hz/m$^2$. When operating at 7\,Volt-over-Voltage (VoV), a 33\% cross talk and an less than 10\% after-pulsing has been identified, well within requirements.
24 SiPMs are placed together on a Printed Circuit Board (PCB) forming the Photo Detection Module (PDM) in an 24\,cm$^2$ area. On the front side of the PCB, the SiPMs are die attached and wire-bonded (Fig.\ref{fig:fig3a}). The readout electronic circuit is populated on the back side of the PCB (Fig.\ref{fig:fig3b}), summing in groups of 4 the response of SiPMs' and perform signal amplification through a transimpedance amplifier (for the TPC) or a custom ASIC (for the neutron veto).  
The Photo Detection Unit (PDU) is comprised of 16 PDMs placed in a 400\,cm$^2$ motherboard (Fig.\ref{fig:fig3c}) that individually biases the PDMs, and performs the signal readout by summing together 4 PDMs. 
The summing scheme (on SiPM, PDM and PDU level) is the optimum compromise between limiting the number of readout channels  and ensuring a sufficiently high signal-to-noise ratio (SNR) despite the 50\,pF/mm$^2$ capacitance, that with only a 5$\times$5\,mm$^2$ surface area passes the $\mathcal{O}(\textrm{nF})$. 
Moreover, to ensure compliance with requirements of radioactivity levels, the prudent selection of materials was followed by extensive screening campaigns using 3 different techniques: ICP-MS, HPGe and $^{210}$Po extraction, covering the upper, middle and lower $^{238}$U chain, and involving four underground laboratories (Boulby, LNGS, LSC, SNOLAB).
\section{Veto  Photodetectors construction and testing}
\begin{figure}[h]
\centering
\subfigure[\label{fig:fig4a}]{\includegraphics[width=0.38\linewidth,height=11pc]{Fig4a.pdf}}
\subfigure[\label{fig:fig4b}]{\includegraphics[width=0.4\linewidth,height=11pc]{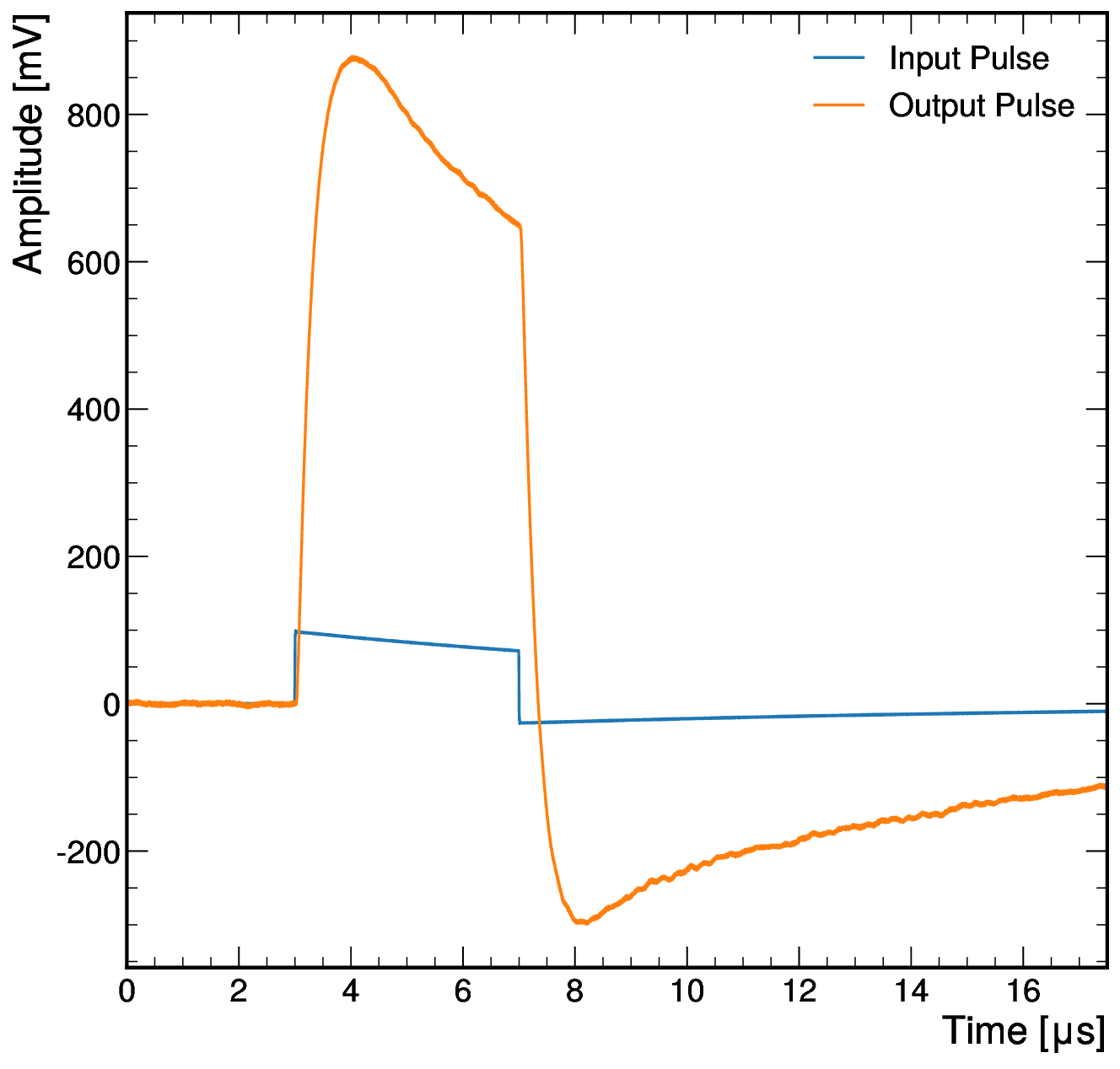}}
\caption{\subref{fig:fig4a} The ASIC interface board. \subref{fig:fig4b} Response of an ASIC to the input test square pulse.\label{fig4}}
\end{figure}
The veto photodetection system comprises of 120 PDUs (vPDU), covering a total area of 5\,m$^2$. The construction, quality assurance and quality control (QA/QC) is performed by several institutes and universities in the United Kingdom. The population of the readout electronics in the veto PDM (vPDM) is performed at the University of Birmingham, the die attach and wire-bonding of the SiPMs at University of Liverpool and STFC-Interconnect. vPDMs are tested in warm and cryogenic conditions at the University of Oxford, whilst the assembly of the vPDU and the warm test is realised at the University of Manchester. Subsequently, the test under cryogenic conditions is performed primarily at the University of Liverpool, whilst smaller facilities have been developed at the University of Edinburgh, University of Lancaster and AstroCeNT (Poland). Additional services are provided by University of Warwick (laser calibration system development) and University of Lancaster (vPDU hospital facility, database). All the operations are performing in ISO5-ISO7 clean room facilities with Rn level <5\,Bq/m$^3$.  At the first step the readout electronics components are populating the Arlon 55NT PCB.
\begin{figure}[h]
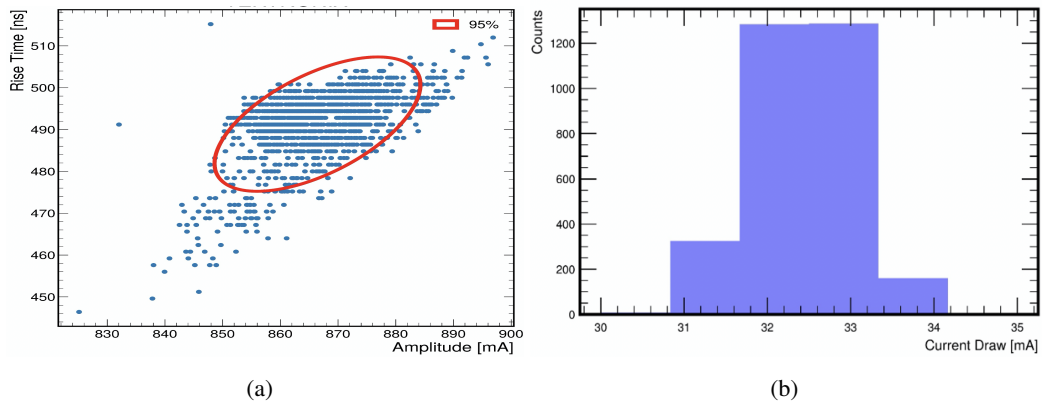

\centering
\subfigure[\label{fig:fig5a}]{\includegraphics[width=0.45\linewidth,height=11pc]{Fig5a.pdf}}
\subfigure[\label{fig:fig5b}]{\includegraphics[width=0.45\linewidth,height=11.2pc]{Fig5b.pdf}}
\caption{\subref{fig:fig5a} Scatter plot of the peak amplitude vs rise-time  and \subref{fig:fig5b} current draw distribution for the total production ASICs.\label{fig5}}
\end{figure}
The performance of the ASIC is tested before soldering to ensure functionality. On this purpose, an ASIC interface board was developed, that emulates the readout electronic circuit (shown in Fig.\ref{fig:fig4a}). The response of the ASIC on a 100\,mV, 4\,$\si{\micro \textrm{s}}$ square pulse (Fig.\ref{fig:fig4b}) and the current draw are recorded. Using a 95\% confidence level ellipse selection on the pulse rise-time vs peak amplitude scatter plot (Fig.\ref{fig:fig5a}) and a 95\% confidence level selection on the Gauss distribution of the current draw, (Fig.\ref{fig:fig5b}) ASICs are selected. The performance of the readout circuit is validated, with the same process, after the pick-and-place procedure and the soldering on the reflow oven.

\begin{figure}[h]
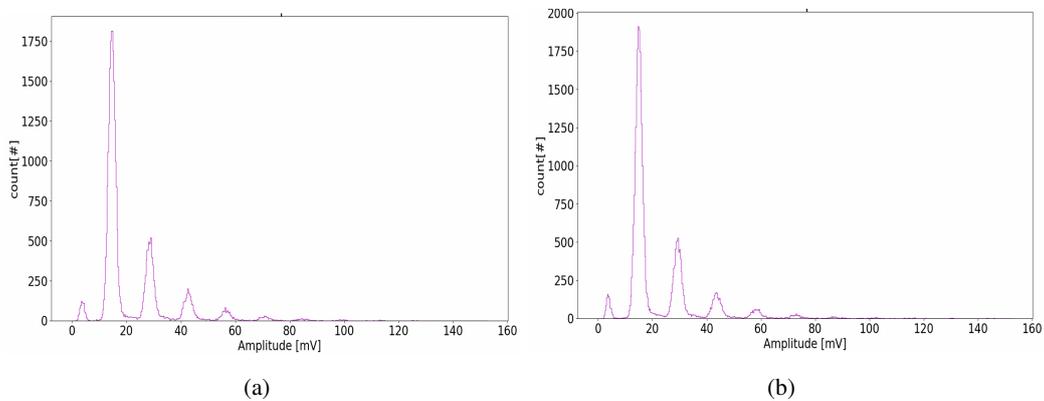

\centering
\subfigure[\label{fig:fig6a}]{\includegraphics[width=0.45\linewidth,height=11pc]{Fig6a.pdf}}
\subfigure[\label{fig:fig6b}]{\includegraphics[width=0.45\linewidth,height=11pc]{Fig6b.pdf}}
\caption{Examples of the response of 2 different quadrants of the same vPDU to the laser calibration pulses.\label{fig:fig6}}
\end{figure}

After die attaching and wire-bonding the SiPMs, the response to single photons for each vPDM is evaluated at cryogenic conditions (liquid N$_2$). 
The criteria include SNR ratio >8 and single photon amplitude between 3.7 and 4.3\,mV when operating on 7\,VoV. The QA/QC criteria for the assembled vPDU have not defined yet due to limited statistics. Fig.\ref{fig:fig6} presents preliminary results on the response of 2 vPDU quadrants (4 vPDMs-summing unit) on the laser calibration pulses. The single photon response found to be $\approx$14\,mV.
\section{Summary}
DarkSide-20k, the largest up-to-date DM experiment using liquid noble gas target, is currently under construction. On the way to an ``instrumental background free experiment'' novel approaches have been adopted, namely: the extraction and distillation of underground Argon,   the neutron veto, and the silicon photodetectors technology. The production of the veto photodetectors has commenced, under intense care for the radiopurity levels, and  statistics are currently building towards effective QA/QC that will validate the expected detector's performance.  
\bibliographystyle{JHEP}
\bibliography{mybibliography}
%
%
%
\end{document}